\documentclass[10pt,letterpaper]{article}
\usepackage{opex3}

\newcommand{\dd}{\ensuremath{\partial}}
\newcommand{\di}{\ensuremath{\mathrm{d}}}

\newcommand{\RabiGTI}{\Omega}
\newcommand{\Rabiin}{\Omega_\mathrm{in}}
\newcommand{\Rabiout}{\Omega_\mathrm{out}}

\newcommand{\aaa}{\ensuremath{U_\Delta}}
\newcommand{\bbb}{\ensuremath{V_\Delta}}
\newcommand{\ccc}{\ensuremath{W_\Delta}}
\newcommand{\ccco}{\ensuremath{W_0}}
\newcommand{\rrr}{\ensuremath{R_\Delta}}

\newcommand{\D}{\mathcal{D}}

\begin{document}

\title{Strong excitation of emitters in an impedance matched cavity:  the area theorem, $\pi$-pulse and self-induced transparency}

\author{Thierry Chaneli\`ere}

\address{Laboratoire Aim\'e Cotton, CNRS UPR3321, Univ. Paris Sud, B\^atiment 505, Campus Universitaire, 91405 Orsay, France}

\email{thierry.chaneliere@u-psud.fr} 



\begin{abstract}
I theoretically study the behavior of strong pulses exciting emitters inside a cavity. The ensemble is supposed to be inhomogeneously broadened and the cavity matched finding application in quantum storage of optical or RF photons. My analysis is based on energy and pulse area conservation rules predicting important distortions for specific areas. It is well supported by numerical simulations. I propose a qualitative interpretation in terms of slow-light. The analogy with the free space situation is remarkable.
\end{abstract}


\bibliographystyle{osajnl}
\bibliography{GTI_bib}

\begin{thebibliography}{10}
\newcommand{\enquote}[1]{``#1''}

\bibitem{haroche2006exploring}
S.~Haroche and J.~Raimond, \emph{Exploring the Quantum: Atoms, Cavities, and
  Photons}, Oxford Graduate Texts (OUP Oxford, 2006).

\bibitem{kozlovsky1988efficient}
W.~J. Kozlovsky, C.~Nabors, and R.~L. Byer, \enquote{Efficient second harmonic
  generation of a diode-laser-pumped {CW Nd: YAG} laser using monolithic {MgO:
  LiNbO}$_3$ external resonant cavities,} Quantum Electronics, IEEE Journal of
  \textbf{24}, 913--919 (1988).

\bibitem{AFCcavity}
M.~Afzelius and C.~Simon, \enquote{Impedance-matched cavity quantum memory,}
  Phys. Rev. A \textbf{82}, 022310 (2010).

\bibitem{PhysRevA.82.022311}
S.~A. Moiseev, S.~N. Andrianov, and F.~F. Gubaidullin, \enquote{Efficient
  multimode quantum memory based on photon echo in an optimal {QED} cavity,}
  Phys. Rev. A \textbf{82}, 022311 (2010).

\bibitem{SabooniCavity}
M.~Sabooni, S.~T. Kometa, A.~Thuresson, S.~Kr{\"o}ll, and L.~Rippe,
  \enquote{Cavity-enhanced storage-preparing for high-efficiency quantum
  memories,} New Journal of Physics \textbf{15}, 035025 (2013).

\bibitem{Schoelkopf.105.140501}
D.~I. Schuster, A.~P. Sears, E.~Ginossar, L.~DiCarlo, L.~Frunzio, J.~J.~L.
  Morton, H.~Wu, G.~A.~D. Briggs, B.~B. Buckley, D.~D. Awschalom, and R.~J.
  Schoelkopf, \enquote{High-cooperativity coupling of electron-spin ensembles
  to superconducting cavities,} Phys. Rev. Lett. \textbf{105}, 140501 (2010).

\bibitem{PhysRevLett.105.140502}
Y.~Kubo, F.~R. Ong, P.~Bertet, D.~Vion, V.~Jacques, D.~Zheng, A.~Dr\'eau, J.-F.
  Roch, A.~Auffeves, F.~Jelezko, J.~Wrachtrup, M.~F. Barthe, P.~Bergonzo, and
  D.~Esteve, \enquote{Strong coupling of a spin ensemble to a superconducting
  resonator,} Phys. Rev. Lett. \textbf{105}, 140502 (2010).

\bibitem{PhysRevLett.107.060502}
R.~Ams\"uss, C.~Koller, T.~N\"obauer, S.~Putz, S.~Rotter, K.~Sandner,
  S.~Schneider, M.~Schramb\"ock, G.~Steinhauser, H.~Ritsch, J.~Schmiedmayer,
  and J.~Majer, \enquote{Cavity {QED} with magnetically coupled collective spin
  states,} Phys. Rev. Lett. \textbf{107}, 060502 (2011).

\bibitem{PhysRevB.84.060501}
P.~Bushev, A.~K. Feofanov, H.~Rotzinger, I.~Protopopov, J.~H. Cole, C.~M.
  Wilson, G.~Fischer, A.~Lukashenko, and A.~V. Ustinov, \enquote{Ultralow-power
  spectroscopy of a rare-earth spin ensemble using a superconducting
  resonator,} Phys. Rev. B \textbf{84}, 060501 (2011).

\bibitem{PhysRevLett.107.220501}
Y.~Kubo, C.~Grezes, A.~Dewes, T.~Umeda, J.~Isoya, H.~Sumiya, N.~Morishita,
  H.~Abe, S.~Onoda, T.~Ohshima, V.~Jacques, A.~Dr\'eau, J.-F. Roch, I.~Diniz,
  A.~Auffeves, D.~Vion, D.~Esteve, and P.~Bertet, \enquote{Hybrid quantum
  circuit with a superconducting qubit coupled to a spin ensemble,} Phys. Rev.
  Lett. \textbf{107}, 220501 (2011).

\bibitem{WilsonNJP}
M.~Afzelius, N.~Sangouard, G.~Johansson, M.~U. Staudt, and C.~M. Wilson,
  \enquote{Proposal for a coherent quantum memory for propagating microwave
  photons,} New Journal of Physics \textbf{15}, 065008 (2013).

\bibitem{JulsgaardPhysRevLett.110.250503}
B.~Julsgaard, C.~Grezes, P.~Bertet, and K.~M\o{}lmer, \enquote{Quantum memory
  for microwave photons in an inhomogeneously broadened spin ensemble,} Phys.
  Rev. Lett. \textbf{110}, 250503 (2013).

\bibitem{van1997dependences}
J.~Van~Wyk, E.~Reynhardt, G.~High, and I.~Kiflawi, \enquote{The dependences of
  {ESR} line widths and spin-spin relaxation times of single nitrogen defects
  on the concentration of nitrogen defects in diamond,} Journal of Physics D:
  Applied Physics \textbf{30}, 1790 (1997).

\bibitem{gao}
W.~Gao, X.-D. Tan, M.-F. Wang, and Y.-Z. Zheng, \enquote{Quantum memory with
  natural inhomogeneous broadening in an optical cavity,} International Journal
  of Theoretical Physics \textbf{52}, 2092--2098 (2013).

\bibitem{PhysRevA.88.012304}
S.~A. Moiseev, \enquote{Off-resonant raman-echo quantum memory for
  inhomogeneously broadened atoms in a cavity,} Phys. Rev. A \textbf{88},
  012304 (2013).

\bibitem{allen1987ora}
L.~Allen and J.~Eberly, \emph{{Optical resonance and two-level atoms}} (Courier
  Dover Publications, 1987).

\bibitem{AreaTheorem}
S.~L. McCall and E.~L. Hahn, \enquote{Self-induced transparency by pulsed
  coherent light,} Phys. Rev. Lett. \textbf{18}, 908--911 (1967).

\bibitem{Moissev_bull}
S.~A. Moiseev, \enquote{Quantum memory for intense light fields in photon echo
  technique,} Izv. Ross. Akad. Nauk, Ser. Fiz. \textbf{68}, 1260 (2004).

\bibitem{Ruggiero2PE}
J.~Ruggiero, J.-L. Le~Gou\"et, C.~Simon, and T.~Chaneli\`ere, \enquote{Why the
  two-pulse photon echo is not a good quantum memory protocol,} Phys. Rev. A
  \textbf{79}, 053851 (2009).

\bibitem{drummond1981optical}
P.~Drummond, \enquote{Optical bistability in a radially varying mode,} Quantum
  Electronics, IEEE Journal of \textbf{17}, 301--306 (1981).

\bibitem{zakharov1995interaction}
S.~Zakharov, \enquote{Interaction of ultrashort light pulses with thin-film
  resonant structures,} Zh. Eksp. Teor. Fiz \textbf{108}, 829--841 (1995).

\bibitem{Zakharov}
V.~A. Goryachev and S.~M. Zakharov, \enquote{Dynamics of transmission of
  ultrashort light pulses by thin-film cavity structures,} Quantum Electronics
  \textbf{27}, 245 (1997).

\bibitem{stenholm1969semiclassical}
S.~Stenholm and W.~E. Lamb~Jr, \enquote{Semiclassical theory of a
  high-intensity laser,} Physical Review \textbf{181}, 618 (1969).

\bibitem{GTI}
F.~Gires and P.~Tournois, \enquote{Interf\`erom\`etre utilisable pour la
  compression d'impulsions lumineuses modul\'ees en fr\'equence,} C. R. Acad.
  Sci. Paris \textbf{258}, 6112--6115 (1964).

\bibitem{pozar2005microwave}
D.~M. Pozar, \enquote{Microwave engineering, 3rd,} Danvers, MA: Wiley  (2005).

\bibitem{Mossberg2001}
C.~Greiner, T.~Wang, T.~Loftus, and T.~W. Mossberg, \enquote{Instability and
  pulse area quantization in accelerated superradiant atom-cavity systems,}
  Phys. Rev. Lett. \textbf{87}, 253602 (2001).

\bibitem{Mossberg2003}
C.~Greiner, B.~Boggs, and T.~W. Mossberg, \enquote{Frustrated pulse-area
  quantization in accelerated superradiant atom-cavity systems,} Phys. Rev. A
  \textbf{67}, 063811 (2003).

\bibitem{CollettPhysRevA.30.1386}
M.~J. Collett and C.~W. Gardiner, \enquote{Squeezing of intracavity and
  traveling-wave light fields produced in parametric amplification,} Phys. Rev.
  A \textbf{30}, 1386--1391 (1984).

\bibitem{GardinerPhysRevA.31.3761}
C.~W. Gardiner and M.~J. Collett, \enquote{Input and output in damped quantum
  systems: Quantum stochastic differential equations and the master equation,}
  Phys. Rev. A \textbf{31}, 3761--3774 (1985).

\bibitem{walls}
D.~F. Walls and G.~J. Milburn, \emph{Quantum Optics} (Springer, 1995).

\bibitem{GorshkovI}
A.~V. Gorshkov, A.~Andr\'e, M.~D. Lukin, and A.~S. S\o{}rensen, \enquote{Photon
  storage in $\lambda$-type optically dense atomic media. {I.} cavity model,}
  Phys. Rev. A \textbf{76}, 033804 (2007).

\bibitem{PhysRevA.85.013844}
B.~Julsgaard and K.~M\o{}lmer, \enquote{Reflectivity and transmissivity of a
  cavity coupled to two-level systems: Coherence properties and the influence
  of phase decay,} Phys. Rev. A \textbf{85}, 013844 (2012).

\bibitem{Eberly:98}
J.~Eberly, \enquote{Area theorem rederived,} Opt. Express \textbf{2}, 173--176
  (1998).

\bibitem{mandel1995}
L.~Mandel and E.~Wolf, \emph{{Optical Coherence and Quantum Optics}} (Cambridge
  University Press, 1995).

\bibitem{Ruggiero:10}
J.~Ruggiero, T.~Chaneli\`ere, and J.-L. Le~Gou\"{e}t, \enquote{Coherent
  response to optical excitation in a strongly absorbing rare-earth ion-doped
  crystal,} J. Opt. Soc. Am. B \textbf{27}, 32--37 (2010).

\bibitem{PhysRevA.58.2733}
L.~Viola and S.~Lloyd, \enquote{Dynamical suppression of decoherence in
  two-state quantum systems,} Phys. Rev. A \textbf{58}, 2733--2744 (1998).

\bibitem{slichter1990principles}
C.~P. Slichter, \emph{Principles of magnetic resonance}, vol.~1 (Springer,
  1990).

\bibitem{PhysRevA.74.033818}
K.~Ichimura and H.~Goto, \enquote{Normal-mode coupling of rare-earth-metal ions
  in a crystal to a macroscopic optical cavity mode,} Phys. Rev. A \textbf{74},
  033818 (2006).

\bibitem{Goto:10}
H.~Goto, S.~Nakamura, and K.~Ichimura, \enquote{Experimental determination of
  intracavity losses of monolithic {Fabry-Perot} cavities made of
  {Pr$^{3+}$:Y$_2$SiO$_5$},} Opt. Express \textbf{18}, 23763--23775 (2010).

\bibitem{sabooni2013three}
M.~Sabooni, Q.~Li, L.~Rippe, and S.~Kr{\"o}ll, \enquote{Three orders of
  magnitude cavity-linewidth narrowing by slow light in a rare-earth-ion-doped
  crystal cavity,} arXiv preprint arXiv:1304.4456  (2013).

\bibitem{Damon}
V.~Damon, M.~Bonarota, A.~Louchet-Chauvet, T.~Chaneli\`ere, and J.-L.
  Le~Gouët, \enquote{Revival of silenced echo and quantum memory for light,}
  New Journal of Physics \textbf{13}, 093031 (2011).

\bibitem{Warren}
F.~C. Spano and W.~S. Warren, \enquote{Preparation of constant-bandwidth total
  inversion, independent of optical density, with phase-modulated laser
  pulses,} Phys. Rev. A \textbf{37}, 1013--1016 (1988).

\end{thebibliography}

\section{Introduction}
Electromagnetic resonators offer the possibility to enhance the interaction of waves with matter. In the RF or optical domain, they are the basis of many commercial devices, going from the microwave-oven to lasers. At the most fundamental level, they permit to explore the boundary between the classical and the quantum world allowing experimental tests of quantum electrodynamics (QED) \cite{haroche2006exploring}.

The practical interest for active resonators, i.e. filled with emitters, has been reactivated in the context of quantum storage for which a complete mapping of the field carrying the information into a long-lived atomic system is necessary. Between the two extreme situations, a single emitter in an ultra-high-finesse cavity (Cavity QED) on one side and a strongly absorbing medium in free-space on the other side, an intermediate regime exists. A weakly absorbing sample can be placed in a medium-finesse cavity to obtain a significant interaction.

I will consider the specific situation of an impedance matched ring cavity. The matching condition means that the input mirror transmission equals the losses. I assume the loss mechanism being dominated by the emitters absorption of the active medium inside the resonator. Conversion into the atomic excitation either way represents a loss or a mapping of the incoming field depending on the point of view. This approach has been particularly fruitful in non-linear optics for second harmonic generation or more generally for frequency mixing \cite{kozlovsky1988efficient}. The incoming power is then fully converted into the target frequency. Following the same idea, the use of a matched cavity has been proposed for quantum light storage \cite{AFCcavity, PhysRevA.82.022311} and successfully implemented in a luminescent crystal \cite{SabooniCavity}. The incoming signal is then fully mapped into the atomic coherences even if the single path absorption is moderate as soon as it is compensated by the cavity quality factor. This situation is not restricted to optics but is also considered very actively in the RF domain with spins \cite{Schoelkopf.105.140501, PhysRevLett.105.140502, PhysRevLett.107.060502, PhysRevB.84.060501, PhysRevLett.107.220501}. Similar proposals have been made to store and retrieve microwave photons \cite{WilsonNJP, JulsgaardPhysRevLett.110.250503}. A resonator, in the optical or in the RF domain, is indeed advantageous as compared to free-space propagation. That's the reason why this technique is actively investigated in a topical context. It allows the use of low absorption samples reducing the constraint on the physical size of the medium or the  concentration of emitters. Highly doped samples can indeed exhibit a important coupling between individual dipoles thus reducing the coherence time. It is the case for example for NV centers in diamond \cite{van1997dependences} and potentially for different kinds of impurities in insulators as rare-earth or transition ions. This statement can be generalized to atomic vapors. Extreme optical depth can be obtained in Bose-Einstein condensate at the price of interaction or collisions between dipoles. Low doping samples in a resonator are clearly an alternative. Using optically thin samples also facilitates the preparation stage as described by Afzelius {\it et al.} \cite{AFCcavity} when optical pumping can be operated from the side of the cavity. In the optical or RF domain, involving optical dipoles or spins, matched resonators offer interesting prospects.

The different protocols \cite{AFCcavity, PhysRevA.82.022311, WilsonNJP, JulsgaardPhysRevLett.110.250503, gao, PhysRevA.88.012304} to store and retrieve quantum signals are based on the coherent manipulation of spins or optical dipoles. A $\pi$-pulse is a tool of choice in that case because it permits the delayed emission of the stored signal in the lineage of the spin or photon echo experiments. The propagation of such strong pulses in free-space absorbing samples is a complex problem known for decades \cite{allen1987ora, AreaTheorem}. More recently, the use of the area theorem has been proposed in the context of quantum memories \cite{Moissev_bull}. This general approach allows the derivation of analytic solutions for weak and strong area pulses. Later on , the distortion though the sample of the rephasing $\pi$-pulse has been identified experimentally as critical to explain the observed storage efficiency \cite{Ruggiero2PE}. My main motivation is to transpose this propagation analysis in a resonator filled with absorbers.

For the rest of the paper, I study the ring cavity design (see fig. \ref{GTI}). The ring configuration has the advantage of the theoretical simplicity. When the round-trip absorption is small, the traveling wave inside interacts only with one atomic mode. As a consequence, inside the resonator, the field and the atomic variables (population and polarization) are fully described by spatially independent parameters (the amplitudes of the field and atomic modes). This model has been extensively and accurately discussed in the context of optical bistability \cite{drummond1981optical, zakharov1995interaction, Zakharov} based on previous work for the semi-classical description of lasers \cite{stenholm1969semiclassical}.

Even if my paper is restricted to the ring resonator design, it should be noted that the standing-wave configuration has the advantage of the experimental simplicity. It only involves two mirrors thus reducing the unwanted passive losses, allowing more compact and stable setup. Semi-monolithic linear resonators are indeed widely used in optics for pulse compression \cite{GTI} or non-linear frequency conversion \cite{kozlovsky1988efficient}. The theoretical treatment is more complex because the standing wave induces a population grating in the non-perturbative regime. It couples the forward and backward propagating modes fundamentally differing from the traveling wave description \cite{drummond1981optical, zakharov1995interaction}. It is beyond the scope of the present paper but it should deserve a specific study because of its experimental simplicity and its current extensive usage for electronic spins coupled to planar waveguides \cite{PhysRevLett.105.140502, PhysRevLett.107.060502, PhysRevB.84.060501, PhysRevLett.107.220501}.

I finally assume the atomic transition to be inhomogeneously broadened (local shift of the transition). It allows the use of photon echo techniques to control the sample and possibly offers a larger intrinsic interaction bandwidth. I assume the coupling constant being identical for all atoms. The assumption is well suited in optics because the orientation of the transition dipoles is well defined with respect to the incoming polarization. It is generally not appropriate for spins in planar waveguides because the magnetic dipoles can have different orientations and the magnetic field orientation around the waveguide has a radial dependency \cite{pozar2005microwave}. Even if I employ the optics terminology and use appropriate assumptions my approach can hopefully open some perspectives for RF or microwave resonators offering a whole bestiary of integrated designs with waveguides on different substrates \cite{pozar2005microwave}.

The aim of this paper is manifold. I primarily study the input/output relations of strong pulses in the matched cavity and point out the possible distortions. I also draw an analogy with the free space propagation through thick sample giving a solid basis for the intuition. I use the intracavity version of the McCall \& Hahn area theorem that has been already derived by Zakharov in the context of optical bistability and superraddiance \cite{zakharov1995interaction, Zakharov, Mossberg2001,Mossberg2003}. The area theorem is extremely helpful to predict the qualitative behavior of strong pulses, $\pi/2$ or $\pi$ for example. $\pi$-pulses for example are fundamental tools for the manipulation of quantum systems. It is still widely unknown in the quantum information community. My goal is to replace the work of Zakharov in this context and show how helpful it is to estimate the possible distortions when the pulse travels through the medium. The analytic part is then essentially a recontextualization of previous results. My numerical calculations of the pulse shapes are strongly supported by the area and energy conservation rules whose simplicity reinforce this kind of analysis. I propose a qualitative interpretation in terms of slow-light. This explanation has the advantage to reintroduce the dispersion properties of the medium particularly useful in the weak signal limit. My approach is clearly inspired by the problematic of optically active emitters in a resonant cavity. Input/output relations are well-known in optics and are formally equivalent to a propagation problem in free space. It is less natural in the microwave domain (mostly in EPR spectroscopy for which relatively large coupling strength can be obtained), because the wavelength is usually larger than the sample. Considering the complete mapping between a microwave or optical photon into spins or optical dipoles respectively brings two fields of study closer. Bridging this gap between optics and RF is also an objective of this paper and is then intended to a larger audience.

\section{Ring cavity model}

\begin{figure}[htbp]
\centering\includegraphics[width=9cm]{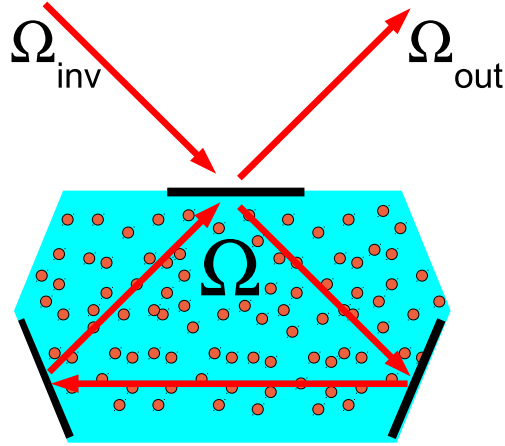}
\caption{Active ring cavity  uniformly filled with emitters. The entrance mirror is partially reflecting. The different fields are defined by their time-varying Rabi frequencies: $ \Rabiin$, $\Rabiout$ represent the incoming and the outgoing amplitudes on the entrance mirror and $\RabiGTI$ the intracavity mode amplitude. The spatial dependence can be neglected when the round-trip absorption is small. The local interference because of a partial beam overlap is supposed to be negligibly within the cavity volume.}
\label{GTI}
\end{figure}

I consider a ring cavity uniformly filled with emitters (fig. \ref{GTI}). In this geometry, the outgoing field $\Rabiout$ and the incoming one $\Rabiin$ can be separated easily. The fields interact with only one mode of the cavity whose amplitude $\RabiGTI$ uniquely characterizes the traveling-wave inside the resonator. All the fields are then completely defined by their time-varying Rabi frequencies.

\subsection{Atoms dynamics and cavity master equations}

The coupled system between the cavity field and the emitters is described on one side by the input/output relations of the resonator including the atomic polarization (source term) and on the other side by the Bloch equation characterizing the dipole dynamics. This model is well-established in quantum optics \cite{CollettPhysRevA.30.1386, GardinerPhysRevA.31.3761, walls, WilsonNJP, JulsgaardPhysRevLett.110.250503} including sophisticated description of multilevel atomic level scheme \cite{GorshkovI} and complete quantum description of QED cavity with inhomogeneous broadening \cite{PhysRevA.82.022311}. Since I specifically study the strong pulse distortions, I use a semi-classical formalism for the cavity master equation and the atomic variables dynamics. It can be naturally extended to describe the interaction at the quantum level \cite{PhysRevA.82.022311,WilsonNJP, JulsgaardPhysRevLett.110.250503,GorshkovI, PhysRevA.85.013844}.

Concerning the dynamics of the emitters, the transition is assumed to be only inhomogeneously broadened. The dipole dephasing and the population lifetime are neglected, in other words the atomic response is examined in the coherent transient regime as in echo type experiments
\cite{allen1987ora}. In the Bloch vector formalism, the transverse and longitudinal decays are then neglected. In the rotating frame the equations read as:

\begin{equation}\label{Bloch}
\begin{array}{rl}
\partial_t  \aaa &= \displaystyle - \Delta \; \bbb \vspace{2mm} \\
\partial_t  \bbb &= \displaystyle\Delta \; \aaa+  \RabiGTI \; \ccc  \vspace{2mm} \\
\partial_t  \ccc &= \displaystyle - \RabiGTI \; \bbb  \vspace{2mm}
\end{array}
\end{equation}
$\Delta$ is the detuning from the excitation frequency and is used as an index accounting for the inhomogeneous broadening. $\aaa$ and $\bbb$ are the in-phase and out-of-phase components of the Bloch vector. The real Rabi  frequency $\RabiGTI$ is the slowly time-varying envelope of the monochromatic field. The last equation describes the population dynamics $\ccc$.

Using the complex amplitude $\displaystyle \rrr= \aaa+ i\bbb$, the dipole moment's equation may be written in a more compact manner:

\begin{equation}\label{P}
\partial_t  \rrr = \displaystyle i \Delta \; \rrr + i  \RabiGTI \; \ccc 
\end{equation}

The cavity master equations also known as input/output relations are slightly modified as compared to the empty ring cavity case \cite{walls}. They include a source term corresponding to the atomic emission:

\begin{equation}\label{BMeqGTI}
\begin{array}{rl}
\displaystyle \frac{1}{\D} \frac{\dd \RabiGTI} {\dd t}& = \displaystyle -\frac{\kappa}{2} \RabiGTI + \sqrt{\kappa} \Rabiin - i\alpha L  \int_{\Delta}  \; g(\Delta) \; \rrr \; \di \Delta \vspace{3mm}  \\
\displaystyle \Rabiout&=\sqrt{\kappa} \RabiGTI - \Rabiin  \\
\end{array}
\end{equation}
The source term is the sum of the polarizations $\displaystyle \rrr$ over the inhomogeneous broadening $g(\Delta)$. I take $\displaystyle g(\Delta) =1$ over the spectrum of interest as the normalization assuming the inhomogeneous broadening is much larger than the interaction bandwidth. It is an important assumption. Even if numerical simulations can be implemented without this condition, it is required by the area theorem as we will see in \ref{intracavity_area}. Since my  paper precisely compares the intuition from the area conservation law and numerical simulations of the temporal shape, I make this assumption from now. To summarize, the dynamics is assumed to verify: homogeneous linewidth $\ll$ interaction bandwidth $\ll$ inhomogeneous broadening \cite[p.3]{Eberly:98}.

It is clearly a restriction because to optimize the resources one would naturally match the bandwidth and the inhomogeneous linewidth as in \cite{PhysRevLett.105.140502, PhysRevLett.107.220501} for example. In that case, the analytic area theorem is not valid anymore, one can only rely on numerical simulations.

Coming back to the model description (eq. \ref{BMeqGTI}), the atoms-light coupling constant is directly included into $\alpha $, the measured absorption coefficient. The convergence of the integral is ensured by $\displaystyle \rrr$ whose support is the excitation bandwidth.

The term $\alpha L$ equals the round-trip absorption of the uniformly filled resonator. It can be rescaled by the filling ratio if the sample is smaller than the cavity.

The cavity is described by the free-spectral range defined as $\D=\displaystyle \frac{c}{L}$ where $L$ is its round-trip length and the intensity transmission coefficient $\kappa$ of the entrance mirror.

I now focus on the specific condition where weak incoming pulses are completely mapped into the ensemble.

\subsection{Matched cavity condition}

The matching condition is defined in the perturbative limit: the populations are not modified (weak pulse) and are supposed to keep its initial value $\ccc = -1$. I define the reflection coefficient in the spectral domain $\displaystyle \widetilde{\RabiGTI}_\mathrm{out}= r\left(\omega\right)\widetilde{\RabiGTI}_\mathrm{in}$ where $\displaystyle \widetilde{\RabiGTI}_\mathrm{out}$ and $\displaystyle \widetilde{\RabiGTI}_\mathrm{in}$ are the Fourier transform of the incoming and outgoing pulses respectively. $r\left(\omega\right)$ completely characterizes the response of the cavity in the weak signal limit. It reads as

\begin{equation}\label{r}
r\left(\omega\right)=\displaystyle \frac{\kappa-2\pi \alpha L - 2i \omega/\D}{\kappa+2\pi \alpha L + 2i \omega/\D}
\end{equation}

The cavity is matched when the reflection is zero on resonance ($ \omega=0$):

\begin{equation}\label{match}
 \alpha L =\frac{\kappa}{2 \pi}
\end{equation}
i.e when the round-trip absorption equals the inverse of the finesse $\displaystyle \frac{2 \pi}{\kappa}$.

For a matched cavity, the intensity reflection $|r\left(\omega\right)|^2$ exhibits a dip whose  FWHM defines the cavity linewidth $\Delta \omega_\mathrm{cav}$. It scales the possible interaction bandwidth as previously demonstrated by Moiseev {\it et al.} \cite{PhysRevA.82.022311, PhysRevA.88.012304}. It is then an important parameter: $\displaystyle \Delta \omega_\mathrm{cav}=2 \kappa \D$.

Because I neglect the atomic decay, the field is completely mapped into the dipoles without damping. It allows the implementation of the field-matter interface. The system is clearly conservative so it is particularly helpfull to use the energy conservation rule to evaluate the conversion from field to atoms.

\section{Energy and area conservation rules}

\subsection{Energy conservation}\label{energyconservation}

In the coherence propagation regime, there is a conservation of quanta between the field (photons) and the excitation of the ensemble (population). Writing the equations by using the Rabi frequencies to describe the field envelope and the absorption coefficient (eqs. \ref{Bloch} and \ref{BMeqGTI}) hides the microscopic parameters (energy quanta) but has the advantage to exhibit experimentally measurable macroscopic parameters. The conservation of quanta can be straightforwardly  retrieved by integrating the equation of motion \cite{allen1987ora}.

On the onde side, the energy contained in the incoming field (electromagnetic) is given by  the integration of the intensity i.e the incoming pulse energy $\displaystyle U_{\Omega} = \displaystyle \int_{t} |\Rabiin|^2 \di t$.

On the other side, the sum of the population gives the energy stored into the atomic excitation. It reads with the same units $\displaystyle U_{w}= \displaystyle \frac{\alpha L}{2 \pi} \displaystyle \int_{\Delta}  \frac{ \ccc +1}{2} \di \Delta$.

Comparing the two forms of energy is a simple way to predict the qualitative shape of the output pulse as we'll see later. This analysis tells us how much energy is potentially left behind by the incoming pulse. The area conservation rule offers a second powerful tool.

\subsection{Intracavity area theorem}\label{intracavity_area}

The McCall-Hahn area theorem is extremely helpful in free space to guide the intuition when propagation in absorbing media is considered \cite{AreaTheorem}. It gives a simple analytic law for the pulse area defined as $\displaystyle \Theta=\int_{t}\RabiGTI\di t$.

Its intracavity version has been derived by Zakharov \cite{zakharov1995interaction, Zakharov}. It has been studied experimentally with doped solids \cite{Mossberg2001,Mossberg2003}. I briefly rederive it here to make the present article more self-contained. I recommend the reading of \cite{Eberly:98} for a pedagogical introduction and accurate derivation to the area theorem in free space (see also \cite[p.19]{allen1987ora} and \cite[p.816]{mandel1995}) . The intracavity version follows the same formal demonstration \cite{zakharov1995interaction}.

It is given by a time integration of the equation of motion (eq. \ref{BMeqGTI}):

\begin{equation}
\begin{array}{rl} 
\frac{\kappa}{2} \Theta - \sqrt{\kappa} \Theta_\mathrm{in} &=\displaystyle - i\alpha L  \int_{t} \int_{\Delta}  \; g(\Delta) \; \rrr(t) \; \di \Delta \di t \vspace{2mm} \\
 \Theta_\mathrm{out}&=  \sqrt{\kappa} \Theta  -\Theta_\mathrm{in} 
\end{array} 
\end{equation}

The source term is resolved by rewriting the Bloch equation (\ref{P}) in its integral form \cite{Eberly:98}. \begin{equation}
\rrr(t) = \displaystyle i\exp(i\Delta t)\int_{-\infty}^t \ccc(t^\prime) {\RabiGTI(t^\prime)}  \exp(-i\Delta t^\prime)\di t^\prime \end{equation}

Following \cite{Eberly:98} because the calculation of the source term contribution is the same as in free space, the distribution $g(\Delta)$ is assumed much broader than the interaction bandwidth so the term $\exp(-i\Delta (t^\prime-t))$ is rapidly oscillating. This simplification is possible because of the three previously mentioned well-separated temporal dynamics: homogeneous linewidth $\ll$ interaction bandwidth $\ll$ inhomogeneous broadening. The integration of the field $\displaystyle \Theta=\int_{t}\RabiGTI\di t$ introduces a time when the field disappears and the macroscopic polarization vanishes because of the inhomogeneous dephasing but the coherences still freely oscillate. This oscillating term is a formal representation of the Dirac distribution centered on $\Delta=0$ \cite{allen1987ora, mandel1995}. After integration over the inhomogeneous profile, only remains $\ccco(t)$, the on-resonance population \cite[eq. (7)]{Eberly:98}. It is the solution of the Rabi flopping problem: \begin{equation}
\ccco(t)=-\displaystyle \cos\left(\int_{-\infty}^{t} \RabiGTI \di t^\prime \right)\end{equation}

I finally obtain the area theorem
\begin{equation}\label{area}
\begin{array}{rl}
\frac{\kappa}{2} \Theta - \sqrt{\kappa} \Theta_\mathrm{in} &= -\pi \alpha L \sin \left( \Theta \right) \vspace{2mm} \\
 \Theta_\mathrm{out}&=  \sqrt{\kappa} \Theta  -\Theta_\mathrm{in}
\end{array}
\end{equation}
It gives input/output relations for the pulse area whatever is the exact incoming temporal shape.

The $\sin \left( \Theta \right)$ term introduces singular propagation effects for specific area integer of $\pi$ as in the free-space situation \cite{Ruggiero:10}. Assuming a matched cavity (eq. \ref{match}) further simply the expression:

\begin{equation}\label{areamatched} \begin{array}{rl} \Theta - \frac{2}{\sqrt{\kappa}} \Theta_\mathrm{in} &= - \sin \left( \Theta \right) \vspace{2mm} \\ \Theta_\mathrm{out}&=  \sqrt{\kappa} \Theta -\Theta_\mathrm{in} \end{array} \end{equation}

\begin{figure}[!h] \begin{center}
\includegraphics[width=10cm]{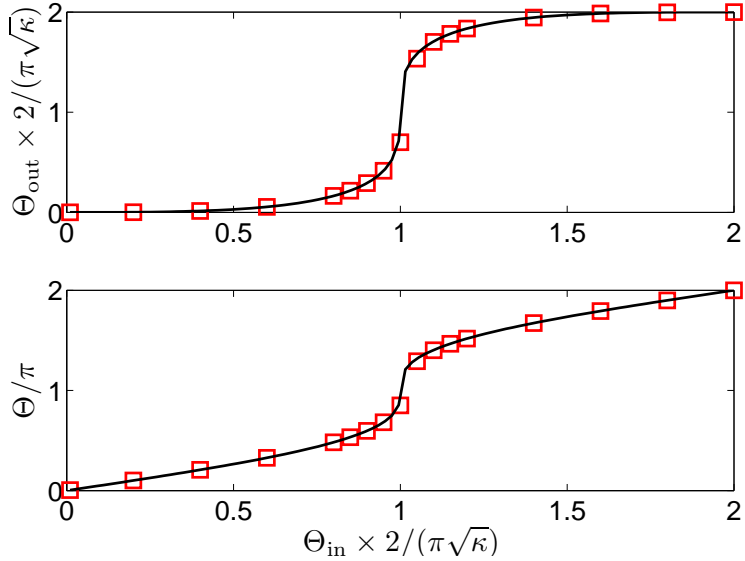}
\caption{Outgoing $\Theta_\mathrm{out}$ (top) and intracavity area $\Theta$ (bottom) as a function of the incoming area $\Theta_\mathrm{in}$ calculated from the matched cavity area theorem  (eqs. \ref{areamatched}).  Squares correspond  to the numerical simulation of the pulse temporal shapes that will be detailed later on (see \ref{simul}). It serves as a validation of the numerically calculated area that can be compared to the analytic result of the area theorem.}
\label{PlotSerie_matchedVSaire_article}\end{center}\end{figure}

In the small area limit (perturbative regime), one verifies that $ \Theta_\mathrm{out} =0$. No light escapes the cavity (matching condition). For incoming area ranging from $0$ to $ \sqrt{\kappa}\pi$, I represent the outgoing and intracavity area (fig. \ref{PlotSerie_matchedVSaire_article}). The latter ranges from $0$ to $ 2 \pi$ accordingly.

I now analyze the two singular situations for the area theorem: $\Theta = \pi$ and $\Theta = 2 \pi$. For both and more generally as any integer of $\pi$, the area is conserved. They are essential for the manipulation of coherences in the context of quantum information storage.

\section{Pulse distortion}\label{distortion}
I here show that energy and area conservation rules can predict the general behavior of strong pulses through the cavity. The intuition is well supported by numerical results.

\subsection{$\pi$-pulses}

\subsubsection{Area and energy of $\pi$-pulses}

I define as $\pi$, the pulse whose intracavity area is $\Theta = \pi$. They correspond to incoming and the outgoing values $\displaystyle \Theta_\mathrm{in}=\Theta_\mathrm{out}= \frac{ \sqrt{\kappa}}{2} \pi$. In other words, the area is conserved. It may be surprising at very first sight because the intracavity $\pi$-pulse is inverting the medium and is then leaving some energy behind. This apparent contradiction actually tells us that the pulse stretches to conserve its area but to reduce its energy. My point is actually to look at the energy conservation to evaluate qualitatively these distortions. If the pulse looses are a neglegible  fraction of its energy, it will keep its shape. It is not the case here. Let's compare the incoming energy and what's left in the atomic excitation by a $\pi$-pulse of bandwidth $\Delta \omega$ (see \ref{energyconservation}). Its amplitude $\Rabiin $ scales as $\Delta \omega \frac{ \sqrt{\kappa}}{2} \pi$ by definition of a $\pi$-pulse. So I obtain $\displaystyle U_{\Omega} \sim \displaystyle \frac{\pi^2}{4} \kappa \Delta \omega $ for the incoming energy. On the other side, if the medium is inverted ($\ccc = 1$) over a bandwidth $\Delta \omega$, the population contains 
$\displaystyle U_{w} \sim \displaystyle \frac{1}{2 \pi} \alpha L  \Delta \omega $.
For a matched cavity (eq. \ref{match}), $U_{\Omega}$ and $U_{w}$ have precisely the same scaling with $\displaystyle \kappa \Delta \omega $. The ratio doesn't depend on the atomic or cavity parameters.

It should be noted that the two forms of energies have the same order of magnitude precisely because the matching condition is assumed. On the contrary, the area theorem in general (eq. \ref{area}) doesn't depend on the absorption for $\Theta = \pi$. In other words, a $\pi$-pulse conserves its area if the cavity is matched or not. It offers a certain degree of freedom if the pulse distortions have to be minimized. If the transmission of the input mirror can be changed for example. Even if the matching is required for a complete storage of the incoming signal, a dynamical control of the resonator parameters as it is the case for RF waveguides would certainly open some perspectives.

\subsubsection{Discussion}

An equivalent energy comparison can be done for an arbitrary area. It tells that the pulses have to leave a constant fraction of energy behind independently from their duration. Let's take weak pulses as a reference. They reduce both area and energy by simply reducing the pulse amplitude keeping the duration constant. This option is not possible for a $\pi$-pulse because the area should be conserved. The only possibility here is to associate amplitude reduction and pulse elongation. A $\pi$-pulse reduces the energy but roughly keeps its area. Major distortions of the incoming pulse are then expected. The analogy with strongly absorbing media in free-space situation is remarkable \cite{Ruggiero:10}.

Before considering numerical simulations which are necessary to obtain the exact shape of the outgoing pulse, I treat the case of Self-Induced Transparency (SIT).

\subsection{Self-induced transparency of $2\pi$-pulses}

The situation is clearly different for $\displaystyle \Theta_\mathrm{in}=  \sqrt{\kappa}\pi$. The area should be still conserved but the energy conservation is relaxed. The on-resonance atoms are returned back to the ground state. Even if off-resonant atoms can still be excited because they don't undergo a $2\pi$ area, less energy should be left behind as compared to a $\pi$-pulse. Because the energy conservation is less drastic, less distortions and a minimized absorption are likely.
Soliton like behavior are expect in analogy with free-space defining the SIT phenomenon \cite{allen1987ora, AreaTheorem}. Since the present paper is primarily focused on quantum storage, $2\pi$ rotations of the atomic state are not really interesting because they essentially leave the system unchanged. $\pi$-pulses are more relevant in that sense. They can indeed rephase the inhomogeneous broadening (photon or spin echo) and be used in series for dynamical decoupling sequences \cite{PhysRevA.58.2733, slichter1990principles}

\subsection{Numerical simulation of the pulse temporal shapes}\label{simul}

\subsubsection{Model and parameters}

The numerical simulation first requires a discretization of the detuning $\Delta$. I write the Bloch system of eq. (\ref{Bloch}) for $n$ evenly-spaced detunings $\Delta_n$ (spacing $\di \Delta$) leading to 3n equations for $(U_{\Delta_n},V_{\Delta_n},W_{\Delta_n})$.

The cavity master (eq. \ref{BMeqGTI}) becomes 

\begin{equation}
\displaystyle \frac{1}{\D} \frac{\dd \RabiGTI} {\dd t} = \displaystyle -\frac{\kappa}{2} \RabiGTI + \sqrt{\kappa} \Rabiin - i\alpha L  \sum_{n}  \; ( U_{\Delta_n}+i V_{\Delta_n}) \; \di \Delta
\end{equation}

It gives $(3n+1)$ linear first-order differential equations for the variables $(U_{\Delta_1},V_{\Delta_1},W_{\Delta_1}, \cdots, U_{\Delta_n},V_{\Delta_n},W_{\Delta_n}, \cdots,\RabiGTI)$.

The boundary conditions are given on the one hand by assuming the atoms in the ground state $(U_{\Delta_n},V_{\Delta_n},W_{\Delta_n})=(0,0,-1)$ and on the other hand by giving an incoming temporal pulse shape $\Rabiin(t)$. I solve the $(3n+1)$ linear system numerically ($n=256$) by applying a forth-order Runge-Kutta method.

For numerical application, I choose a finesse of $\displaystyle \frac{2 \pi}{\kappa}=500$ and a free-spectral-range $\D= 3 \mathrm{GHz}$. These common parameters covers different physical realities. In the optical domain, they correspond to a 10 cm long cavity for which finesse of  1000 have been observed \cite{PhysRevA.74.033818, Goto:10} using monolithic samples. In the RF domain, a GHz range can be associated with the hyperfine splitting of NV centers in diamond \cite{PhysRevLett.105.140502, PhysRevLett.107.060502, PhysRevLett.107.220501} or with the electron spin of paramagnetic impurities in luminescent crystals \cite{Schoelkopf.105.140501, PhysRevB.84.060501}. The finesse in that case is not limited by the ultimate performance of the superconducting resonator but by the atomic linewidth (typical in the MHz range) then limiting the quality factor (few hundreds). A finesse of $\displaystyle \frac{2 \pi}{\kappa}=500$ is then realistic and covers a wide range of physical systems.

The cavity is supposed to be matched (eq \ref{match}). The different parameters are now fixed.

\subsubsection{Outgoing pulse shapes}

\begin{figure}[!h] \begin{center}
\includegraphics[width=11cm]{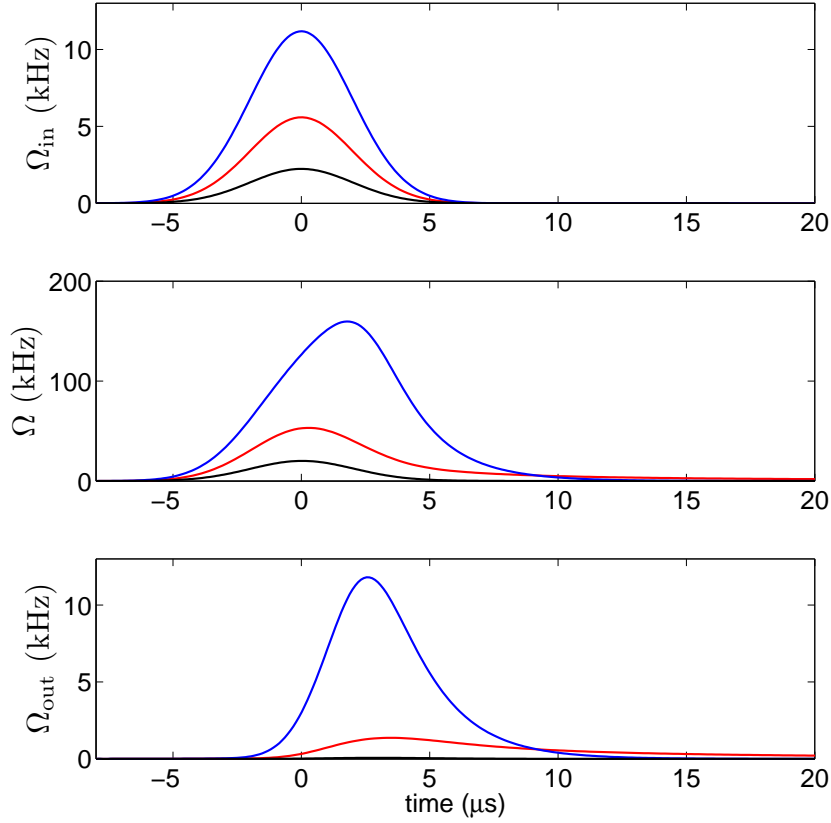}
\caption{Top: incoming gaussian pulses $ \Rabiin$ of varying area $\Theta_\mathrm{in}= 0.4 \frac{ \sqrt{\kappa}}{2} \pi$ in black, $\Theta_\mathrm{in}=  \frac{ \sqrt{\kappa}}{2} \pi$ in red ($\pi$-pulse) and $\Theta_\mathrm{in}=  \sqrt{\kappa} \pi$ in blue ($2\pi$-pulse). The corresponding intracavity $\RabiGTI$ (middle) and outgoing pulses $\Rabiout$ (bottom).}
\label{PlotSerie_matchedTemporal_article}\end{center}\end{figure}

I choose gaussian shaped incoming pulses for the simulation with a $2\mu s$ duration. The corresponding bandwidth is $2 \pi \times 80\mathrm{kHz}$ much smaller than the cavity linewidth $\Delta \omega_\mathrm{cav}=2 \pi \times 12 \mathrm{MHz}$. I keep the duration constant and vary the pulse amplitude ranging the incoming area from zero to $\Theta_\mathrm{in}= \sqrt{\kappa} \pi$ ($2\pi$-pulse). I then obtain the temporal shape of the intracavity and outgoing pulses in fig. \ref{PlotSerie_matchedTemporal_article}.

I also perform a rudimentary test of my simulation by computing the intracavity $\Theta$ and outgoing $\Theta_\mathrm{out}$ areas. They can be easily compared (squares on fig. \ref{PlotSerie_matchedVSaire_article}) to the analytic result of the area theorem (eq. \ref{areamatched}).

The pulse for $\Theta_\mathrm{in}= 0.4 \frac{ \sqrt{\kappa}}{2} \pi$ behaves as a weak area pulse meaning it is mostly absorbed without distortion. No reflection at all is expected for areas much smaller than $\frac{ \sqrt{\kappa}}{2} \pi$.

The $2\pi$-pulse with $\Theta_\mathrm{in}=  \sqrt{\kappa} \pi$ shows remarkable similarities with SIT in strongly absorbing media. It conserves its amplitude and shape. It is essentially delayed. The delay scales as the incoming pulse duration, $2\mu s$ in that case \cite{allen1987ora}. The situation is particular and has been observed experimentally  \cite{Mossberg2001, Mossberg2003}. It is less interesting in the context of quantum manipulation of qubits but it would certainly deserve further investigation. I now consider more specifically the $\pi$-pulse situation.

\begin{figure}[!h] \begin{center}
\includegraphics[width=12cm]{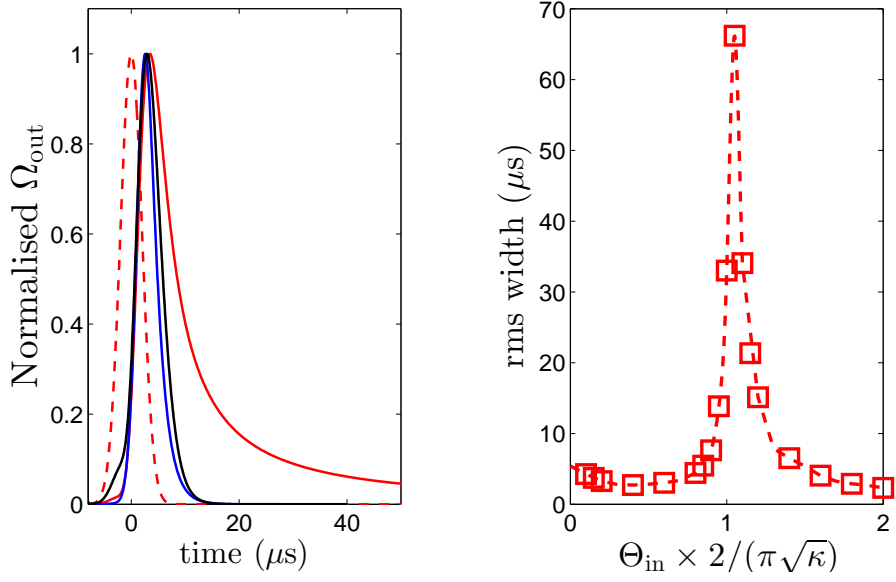}
\caption{Left: normalized outgoing pulse $\Rabiout$ for $\Theta_\mathrm{in}= \displaystyle 0.4 \frac{ \sqrt{\kappa}}{2} \pi, \frac{ \sqrt{\kappa}}{2} \pi$ and $ \displaystyle \sqrt{\kappa} \pi$ in black, red and blue respectively (as in fig. \ref{PlotSerie_matchedTemporal_article}). The normalised incoming pulse is plotted as a reference (dashed red). The long tail of the outgoing $\pi$-pulse (red) is prominent. Right: root mean square (rms) temporal widths of the calculated pulse (squares, a dashed line is used to guide the eye).}
\label{PlotRMSwidthVSaire_article}\end{center}\end{figure}

The outgoing shape  $\Rabiout$  for $\Theta_\mathrm{in}=  \frac{ \sqrt{\kappa}}{2} \pi$ is characterized by a long tail, much longer than the pulse duration. This strong distortion is precisely expected for $\pi$-pulse which have to leave most of its energy behind by conserving its area (see section \ref{distortion}). To make this comparison more clear, I plot the normalized outgoing pulse (fig. \ref{PlotRMSwidthVSaire_article}, left).

As previously mentioned, the $2\pi$ outgoing pulse (SIT) is remarkably similar to the incoming pulse (fig.  \ref{PlotRMSwidthVSaire_article}, left, blue curve). It is essentially delayed. The almost weak area pulse $\Theta_\mathrm{in}= 0.4 \frac{ \sqrt{\kappa}}{2} \pi$ (fig.  \ref{PlotRMSwidthVSaire_article}, left, black curve) is not distorted neither. For a matched cavity, it is strongly absorbed, the observed shape and delay should be taken with precaution because they strongly depend on the numerical parameters. Temporal discretization may typically explain the pedestal on the rising edge of outgoing pulse (fig.  \ref{PlotRMSwidthVSaire_article}, black curve). Anyway, it serves as a reference as compared to $\pi$-pulse whose behavior is completely different. A long tail clearly appears.

\subsubsection{Outgoing pulse width}

To quantify the distortion, I plot the rms width $\sigma$ of the pulse (fig. \ref{PlotRMSwidthVSaire_article}, right). This standard deviation is computed from the normalized distribution $p_\mathrm{out}(t)=\displaystyle \frac{\Rabiout(t)}{\Theta_\mathrm{out}}$ and defined as $\sigma = \sqrt{ \displaystyle \int_t \left( t - \mu \right)^2 p_\mathrm{out}\, \di t}$ where $\mu$ is the mean value $\mu= \displaystyle \int_t  t p_\mathrm{out}\, \di t$. It equals $2\mu s$ for the incoming pulse as reference. For incoming area close to $\Theta_\mathrm{in}= \frac{ \sqrt{\kappa}}{2} \pi$ elongations by more than an order of magnitude are observed. The distortion is extremely sensitive to the pulse area close to $\frac{ \sqrt{\kappa}}{2} \pi$. The numerical simulation allows to put numbers on my qualitative analysis based on area and energy conservation.

\subsubsection{Slow-light interpretation}\label{SL}

The pulse distortion is usually associated with a steep dispersion profile. The slow-light phenomenon recently reappeared in the context of impedance matched cavity \cite{sabooni2013three} with particularly strong and observable effects. Considering the reflection coefficient (eq. \ref{r}) as an effective susceptibility whose imaginary part is the equivalent refractive index is tempting. It would be only justify in the weak signal perturbative limit and cannot be used in general to explain the outgoing shape of strong area pulses. I nevertheless propose to consider the effective group delay for a qualitative analysis essentially demonstrating that extreme dispersion is expected for an inverted medium. One can push the perturbative approach by looking at the reflection coefficient  (eq. \ref{r}). I assume for example that a given time the atomic population is excited $\ccc(t) \geq -1$, and from that point calculate an effective reflection coefficient. I now assume the population constant $W$ over the bandwidth of interest for the sake of simplicity since my goal is to derive an order of magnitude. The population simply rescales the absorption coefficient so to say taking the form $-W \times \alpha$. It leads to gain or absorption if the medium is inverted or not. Starting from the generalized expression of the reflection coefficient

\begin{equation}\label{r_gen}
r_W\left(\omega\right)=\displaystyle \frac{\kappa+4\pi W \alpha L - 4i\pi \omega/\D}{\kappa-4\pi W \alpha L + 4i\pi \omega/\D}
\end{equation}

One can derive the group delay $T_g$ for a matched cavity (eq. \ref{match}) by a first order expansion

$ \displaystyle
r_W\left(\omega\right)=r_W\left(0\right)+i\omega T_g+...
$

One simply finds 

\begin{equation}\label{T_g}
r_W\left(0\right)=\frac{1+W}{1-W}\mbox{\hspace{1cm} and  \hspace{1cm}} T_g=\frac{1}{\Delta \omega_\mathrm{cav}}\frac{4}{\left(1-W\right)^2}
\end{equation}

I now compare the initial situation in the ground state (the population is weakly modified by small area pulses) and a inverted medium as expected for $\pi$-pulses. For atoms in the ground state, the reflection tends toward zero and the group delay toward $\displaystyle \frac{1}{\Delta \omega_\mathrm{cav}}$, that can be compared to the empty ring cavity case with a $\displaystyle \frac{4}{\Delta \omega_\mathrm{cav}}$ delay. When the medium is inverted, both the reflection $r_W\left(0\right)$ and the group delay diverge. It is extremely surprising at first sight because an infinite reflection seems to break the energy conservation rule. It is not the case because it is precisely associated with an infinite group delay. It tells that the weak pulse stays for an infinitely long time in the cavity. Because the ensemble is supposed inverted (and not depleted), it gives constantly its energy to the pulse leading to an endless amplification $r_W\left(0\right) \rightarrow \infty$. This analysis should not be confused with the distortion of a $\pi$-pulse. A strong pulse enters an not-inverted medium but creates the inversion through which it propagates. It cannot be accurately described by my perturbative expansion and precisely requires a numerical simulation to solve the coupled equations. We nevertheless understand qualitatively that the more the pulse inverts the medium, the slowest is goes through the sample. It is not so surprising at the end that a long tail appears following the outgoing $\pi$-pulse.

\section{Conclusion}
The intracavity and output shapes of a strong exciting pulse are extremely sensitive to its incoming area close to $\Theta_\mathrm{in}= \frac{ \sqrt{\kappa}}{2} \pi$. It corresponds to a singularity for the intracavity area theorem. This effect can be interpreted as a competition between the area and energy conservation rules. The pulse stretches because it reduces its energy by keeping its area. The parallel with the free space propagation is remarkable. This effect appears to be critical to explain the observed efficiency of the two-pulse echo in an optically-thick sample 
\cite{Ruggiero2PE}. A similar result is expected for echo type experiments in a matched cavity.

A long pulse tail can be problematic when a weak signal has to be isolated from a strong control pulse. To get around this potential limitation, monochromatic $\pi$-pulses can be replaced advantageously by adiabatic inversion as proposed in free-space \cite{Damon}. Frequency swept pulses are not only more robust in terms of power fluctuation for example but they are known to undergo less distortion when propagating though absorbing samples \cite{Warren}. A similar behavior is expected in a matched-cavity. Frequency swept pulses would deserves a specific study.  The Bloch-Maxwell model can be extended to include the time-dependent phase of the field in a straightforward manner \cite{allen1987ora}. It should be noted that the area theorem is not valid anymore, in a certain sense relaxing the constraints on the pulse dynamics \cite{Eberly:98}.

\section*{Acknowledgments}
I would like to acknowledge useful discussions with P. Bertet, J.-L. Le Gou{\"e}t and A. Ourjoumtsev as well as financial support from the french national grant RAMACO no. ANR-12-BS08-0015-02. The research leading to these results has received funding from the People Programme (Marie Curie Actions) of the European Union's Seventh Framework Programme FP7/2007-2013/ under REA grant agreement no. 287252.

\end{document}